# Energy trapping and shock disintegration in a composite granular medium

C. Daraio[1], V.F. Nesterenko[1,2*], E.B. Herbold[2], S. Jin[1,2]

[1]*Materials Science and Engineering Program*

[2]*Department of Mechanical and Aerospace Engineering*

*University of California at San Diego, La Jolla, CA 92093-0411, USA*

ABSTRACT

Granular materials demonstrate a strongly nonlinear behavior influencing the wave propagation in the medium. We report the first experimental observation of impulse energy confinement and the resultant disintegration of shock and solitary waves. The medium consists of alternating ensembles of high-modulus vs orders of magnitude lower-modulus chains of spheres of different masses. The trapped energy is contained within the "softer" portions of the composite chain and is slowly released in the form of weak, separated pulses over an extended period of time. This effect is enhanced by using a specific group assembly and a superimposed force.

Strongly nonlinear systems, e.g. one-dimensional chains of beads, exhibit a very unique wave behavior [1], especially at the interface between two different granular systems [1-6] or at the interface of granular media and solid matter [7].

Granular beds composed of iron shot (the waste from metallurgical plants) have been successfully used as shock-mitigating protectors in the design of explosive chambers that reduce the amplitude of shock waves generated by a contact explosion [1]. The understanding of their fundamental behavior may help to create a new class of functional composite materials with novel dynamic properties. In the past, the design of shock protectors focused mainly on the wave transformation provided by layered systems or on the enhanced energy dissipation in porous media [1,8,9]. Yet another, entirely different way of protecting materials is through the confinement of an impulse in a particular region of the shielding medium called a *"granular container"* as predicted by theoretical analysis [3,4], but has not been experimentally demonstrated.

Granular matter, common in our everyday life, has many known applications but it presents fundamental difficulties in the understanding of its intrinsic dynamic properties due to the strong nonlinearity and complex contact-force distributions between grains [1,10-12]. Their three dimensional structural features include filamentary force chains which may be relevant to characterize the behavior of other matters such as in a glassy state [11] and in our case extend our results to ultra-short pulse propagation in this matter.

The study of the one-dimensional case is therefore of general interest for fundamental understanding of the strongly nonlinear behavior of these complex media. Practical applications can subsequently arise extending, for example, the properties of the 1-D system to a three dimensional array of ordered granular chains. In this Letter we describe the first experimental discovery of forced confinement of propagating impulses (solitary and shock waves) inside a composite granular medium. The one dimensional model system is used to experimentally demonstrate the concept of impulse trapping proposed in [3,4] for incident solitary wave pulses. A system with similar dynamic behavior can be made from other different structural elements with strongly nonlinear interactions.

The strongly nonlinear behavior in a chain of elastic spherical beads arises from the nonlinearity of the Hertzian contact interactions between the particles composing the system and results in a power-law type dependence of the compressive force ($F$) on displacement ($\delta$) (where $F \propto \delta^{3/2}$) combined with zero tensile strength. In the case of zero or very weak precompression (resulting in zero or very small sound speed, i.e. "sonic vacuum" SV type systems) the corresponding wave equation supports a qualitatively new solitary wave [1]. A peculiar property of the granular media derives from the possibility of "tuning" the type of stationary solution produced by the system by varying the precompression acting on the chains [1,15,16]. This allows "choosing" the regime of wave propagation or the reflection from the interfaces of two SVs according to the needs for each specific application.

The passage of a solitary wave through the interface of two SV type systems from a region of particles with a higher elastic modulus (or higher mass) to a region of lower elastic modulus (or lower mass) results in the impulse decomposition into a train of solitary pulses. In the zero or weakly precompressed case, the number of pulses composing the train is proportional to the ratio of the difference in the mass of the particles at the two sides of the interface. In this case, no reflected wave from the interface is observed propagating back into the stiffer region. In contrast, when the solitary wave in SV passes from the softer (lower elastic modulus) region to the stiffer region, it divides its energy into 2 portions. In this case, no impulse disintegration beyond the interface is observed [1-5]. The reflection from light and heavy inclusions was earlier proposed as a technique for nondestructive identification of impurities in a granular medium (with implications in the analysis/detection in geological or biological fields) [17]. No numerical or experimental results were published on the trapping of shock type impulses which are qualitatively different from solitary waves and most important for practical applications.

To create the "granular container" for pulse trapping, we used a total 32 beads, of which 22 beads were the high-modulus, large mass stainless steel beads (non-magnetic, 316 type) and 10 were the low-modulus, small mass PTFE (polytetrafluoroethylene) beads which support strongly nonlinear solitary waves [18]. The diameter of the beads was uniform, ~4.76 mm, and the bead arrangements in different configurations were investigated. The mass of a 316 stainless steel bead was 0.45 g, with a density of 8000 kg/m$^3$, Young's Modulus of 193 GPa and the Poisson ratio equal to 0.3 [19,20]. The mass of a PTFE bead was 0.123 g, the density 2200 kg/m$^3$, the elastic modulus was 1.46 GPa, and a Poisson ratio 0.46 [21,22]. They were chosen to demonstrate that the pulse trapping effect is sensitive to the geometrical arrangement of the same number of structural elements for the enhanced protection of the bottom wall from the incoming impulse. Three piezo-sensors were embedded inside particles in the system as described in [5,16,18] allowing the time-of-flight calculations of the pulse speed. A fourth sensor was embedded in the wall at the bottom of the chain. The particles were assembled in a vertical PTFE holder. Pulses were generated with a 0.47 g $Al_2O_3$ rod striker dropped from various heights for the single solitary wave type loading and also with a much heavier, 63 g $Al_2O_3$ rod for the shock-type loading. In order to tune the properties of this new "granular protector", a magnetically induced non-contact compressive force (2.38 N) was applied as in [5,16].

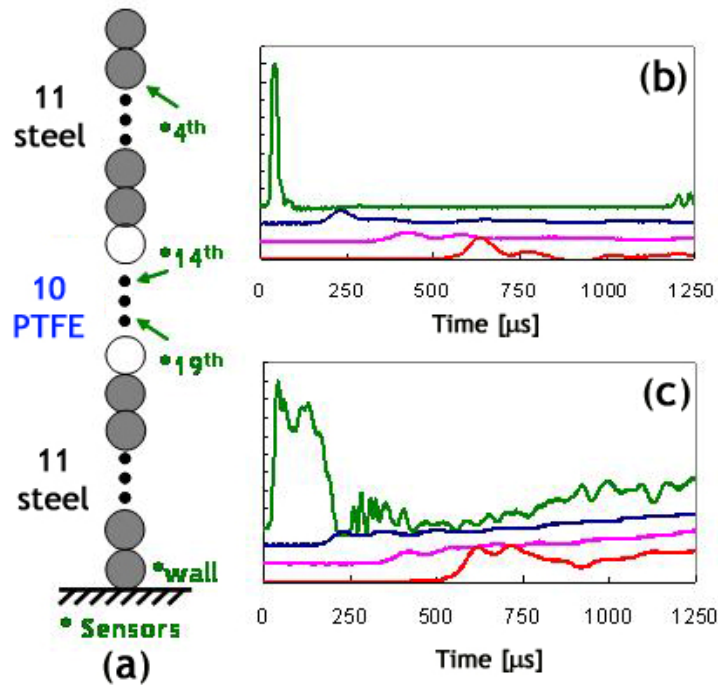

FIG. 1 (Color Online) Trapping of a solitary- and a shock-pulse in the composite single "granular container". (a) Schematic diagram of the stainless steel and PTFE beads geometrical arrangement used for testing with indicated sensors. (b) Experimental data corresponding to the solitary-type loading. (c) Experimental data corresponding to the shock-type loading. The y-axes scale for all curves is 1 N per division.

First, a *"single granular container"* was tested (Fig. 1(a)), in which 11 stainless steel beads were placed at the top of the chain, 10 PTFE beads in the middle and 11 steel beads at the bottom, forming a softer central section of the chain. The corresponding impulse behavior is presented in Fig. 1(b,c) for incident solitary and shock waves.

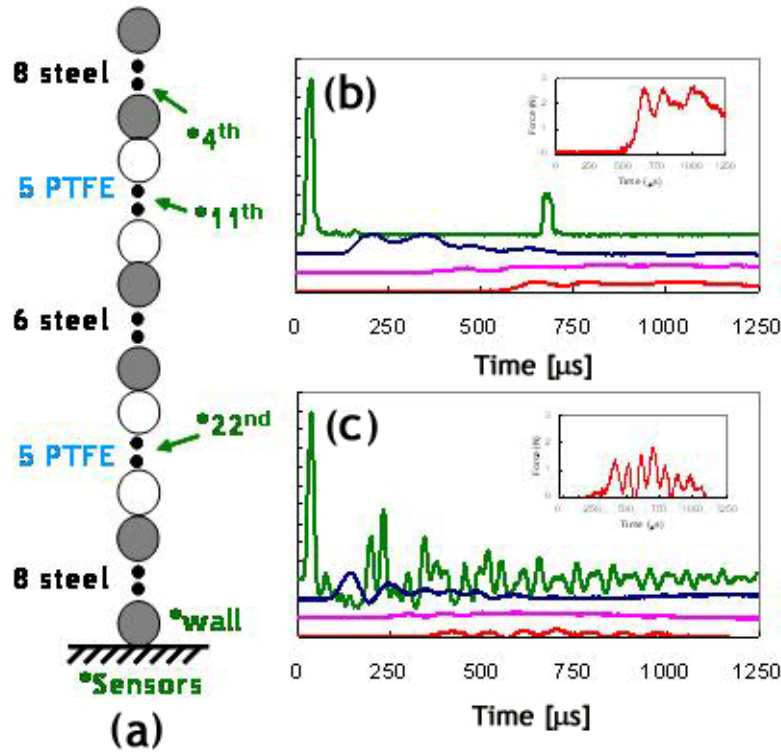

FIG. 2 (Color Online) Solitary pulse trapping in the composite double "granular container" with and without additional precompresion. (a) Schematic diagram of the arrangement of the stainless steel and PTFE beads with indicated sensors. (b) Experimental results for only gravitationally loaded system. (c) Experimental results with magnetically induced superimposed force, all other conditions as in (b). The y-axes scale is 1 N per division. Insets in (b,c) show the pulse behavior at the wall for the gravitationally loaded and the magnetically tuned system. Note the significantly moderated impulse shape arriving on the wall in (b) and (c): the strong incident impulse (first curve) disintegrates into a very weak series of pulses delivered over a much longer period of time (bottom curve).

The *"double granular container"* (Fig. 2(a)) configuration consisted of the same overall number of stainless steel and PTFE beads, but they were divided into two 5-bead PTFE sections interposed between the stainless steel beads.

Numerical analysis of the discrete chains was performed for all the set-ups described for the calculation of force-time curves as well as for the total energy trapped and released by the "granular containers". The numerical simulations were run similar to [16,18] using the equation of motion for the grains with Hertzian contact [1]. The presence of the gravitational precompression (caused by the vertical orientation of the tested chain) was taken into account in the numerical analysis although it has a weak effect at the investigated pulse amplitudes. The effects of dissipation were not included in the calculations and will be addressed in future studies.

Figure 1(b,c) shows the experimental results obtained in the *"single granular container"*. In this case, the trapping of the incident solitary pulse in the softer region is clearly evident in experiments (Fig. 1(b)) and was well matched in the numerical calculations. The experimental data clearly demonstrates that the incident solitary pulse (~40 μs long and 8 N in amplitude) is quickly transformed by the PTFE portion of the chain to a much longer signal and it is decomposed into a train of pulses arriving at the wall with a much lower amplitude.

The solitary wave speed in the steel section of the chain is 357 m/s. The experimental data (Fig. 1(b)) shows that the signal speed drops down to 137.4 m/s when the pulse passes through the first interface into the PTFE section. This slowing down, caused by the drastically lower elastic modulus of PTFE (1.46 GPa) as compared to steel (193 GPa), enables the pulses to remain mostly trapped in the softer section of the chain for a relatively long time, "bouncing" back and forth between the two stiffer groups (top and bottom interfaces), releasing the energy of the impact in both directions very slowly.

The first and largest impulse reaching the wall in Fig. 1(b) has an experimentally measured amplitude significantly smaller (about 5 times) than the one measured in independent experiments in a uniform steel chain under identical impact conditions and the same number of particles. In the numerical calculations the reduction is also very significant being 3 times smaller (amplitude 6.7 N) than the wall amplitude of ~20 N in uniform stainless steel beads chain. The difference between experimental data and numerical results is most likely due to the absence of dissipation assumed in the numerical calculation, which underestimates the total extent of signal amplitude reduction. It also demonstrates that the dissipation present in the experiments can significantly enhance the protection against incident solitary waves. The trapped pulses reflected from the bottom of the soft section have an amplitude comparable with the incoming pulses in the PTFE section, demonstrating that a significant amount of the total energy remains confined in the softer central portion of the chain, slowly leaking out only a small amount at each rebound (~28% after the first rebound).

Performance of the *"single granular container"* against shock wave type loading achieved in experiments is shown in Fig. 1(c). To the best of our knowledge its trapping in a *"granular container"* was not investigated before theoretically or experimentally, though shock loading is very important in practical applications. A shock wave is a qualitatively different type of pulse in comparison with solitary wave. It is usually characterized by a longer duration which may affect the reflection and transmission at the interfaces. The results of numerical calculations for both types of incident waves (solitary and shock), showed similar tendencies in the impulse behavior in qualitative agreement with the experimental results.

To investigate the influence of particle arrangement in our system and to enhance the protection efficiency, we reorganized the PTFE beads as shown in Fig. 2(a). Here we alternated steel and PTFE portion of the chains with a periodicity 8-5-6-5-8, where the 5 particle portions are composed of PTFE beads only. This *"double granular container"* arrangement had the same number of stainless steel and PTFE beads in the two segments of the soft section, in order to enable a more direct comparison of the wall protection. Figure 2(b) shows the experimental results with demonstrate a significant improvement in the trapping of a single solitary wave pulse in the "*double granular container"*.

This geometry resulted in a much better protection of the bottom wall by more efficiently trapping most of the incoming pulse (compare Fig. 2(b) with Fig. 1(b)) and releasing its energy more slowly. In this case the amplitude of the force analyzed numerically at the wall was 3.4 times less (2.4 in experiments) than the one detected in the "*single granular container*" and ~10 times less than the one observed in an all-steel chain. This discrepancy between the numerical and the experimental case is probably due to the enhanced effects of dissipation at higher signal amplitudes. However, there exists a qualitative agreement of the wave behavior in experiments and numerical calculation for solitary-type loading.

In this set-up, the first (uppermost) section of the PTFE works very efficiently trapping a larger amplitude of the pulse and transforming the 40 μs long incoming solitary pulse (from the steel section) into a much longer and delayed train of signals with an overall duration over 1000 μs. Numerical calculations of the energy constrained in the "granular container" confirmed the higher efficiency as a protector: the "*double container*" traps most of the potential energy for a longer time when compared to the single "granular container" configuration.

It was previously reported that the wave behavior and the reflection from the interface of two strongly nonlinear systems is strongly affected by an initial precompression [5,16]. This effect might be used to control and improve the protecting behavior of the investigated systems. To explore the influence of the superimposed forces we tested the more efficient "*double granular container*" under a magnetically induced precompression. This resulted in an evident increase of the speed of the signal propagation and in the creation of an anomalous reflected wave [5] on the first steel sensor (uppermost curve) followed by a series of multiple reflected pulses (compare Figs. 2(b) and 2(c)). The introduction of the preload significantly reduced the force impulse acting on the wall, facilitating the splitting of the signal into a train of low-amplitude waves (see insets in Fig. 2(b) and 2(c) showing in details the pulse behavior at the wall).

The physical explanation for such an increase in the pulse confinement in the softer region of the chain is related to the self assembly of gaps at the interfaces causing a complex "rattling" among the interfacial particles combined with the reflection of the pulse from the interfaces of the soft and rigid regions. These gaps allow the two "granular containers" to keep the energy trapped longer, therefore enhancing the protection of the wall. Moreover when the signal propagates through the first interface, a "fracture wave" is formed and propagates back into the stainless steel chain. The presence of these opening and closing gaps is counter intuitively enhanced by the static precompression and it is responsible for the introduction of a new time-scale in the system as well as the formation of an anomalous reflected wave at the interface under precompression (top curve of Fig. 2(c)). As a result, the gaps delay the wave reflection and propagation, and enhance the backward reflections from the heavy/light interfaces. In this case the total energy trapped in the softer sections remains almost constant within the investigated time. Furthermore the superimposed force transforms the pulse arriving at the wall in a series of well separated impulses, reducing the total momentum reaching the bottom wall. This behavior is very useful as a mean to protect an object from incoming impacts by providing longer distances of pulse traveling within the protector region, thus additionally causing the impact to lose its energy due to dissipation.

The "*double granular container*" was also tested for trapping of shock pulses. To generate such pulses we used an $Al_2O_3$ rod (63 g) as a striker impacting the first steel bead. The signal reaching the wall was dramatically transformed from an oscillatory, fast-ramping shock loading into a long, slowly increasing series of pulses. This so trapped and transformed pulse is likely to be much less damaging to the protected object (the end wall in this experiments).

Results of the numerical calculations indicated a similar trend as in the experiments. Interestingly the data demonstrates that under shock-type loading the softer sections of the chain do not appear to trap energy, thus only acting as pulse transformers, as opposed to the energy trapping of incident solitary waves.

Calculations were also performed for a chain composed of one-by-one alternating stainless steel and PTFE beads, to see if the increasing the number of contacts between the two different beads throughout the chain further improves the shock protection. In this case the chain responded as a homogenized "two-particle system" [1] without creation of reflected pulses, thus drastically reducing the efficiency of the protector.

The shock-disintegrating principles demonstrated here can be utilized for practical composite structures. 3-dimensional composite materials can be prepared using various approaches, such as a modified or multilayer version of aligned composite media [13,23] the results of which will be reported in future publications. Among the foreseen practical application for such a novel, pulse-disintegrating material are protection gears for military or construction hazards, sound-proof coatings or layers for buildings, protection devices for human ear and other parts on possible exposure to extreme pulses or explosions, and soft-landing of spacecrafts or highly protective shipping box for delicate machineries. Biomedical applications are also conceivable, for example, as a pulse-preventing layer for sensitive regions of human body where the acoustic beam is to be avoided during ultrasonic treatment of brain tumors or deposited stones.

In conclusion, we presented a new and efficient pulse-disintegration experimental phenomenon in specially designed composite materials. We demonstrated experimentally and numerically the efficiency of soliton-like and shock-like pulse trapping and slowed-down energy leaking in a high-/low-modulus (mass) composite structured "granular container" and proved that the efficiency of the protector depends on the nature of the particle arrangements. Under shock-type loading a drastic modification of the signal ramp time at the wall was obtained using a *"double container"* configuration. The introduction of a magnetically induced precompression divided the signal reaching the wall into a series of subdivided pulses reducing the total force impulse even further. If properly configured, these grouped composite media can be building blocks for powerful energy absorbers against impacts, and can be useful as efficient protectors for technological and security applications.

The authors wish to acknowledge the support of this work by the US NSF (Grant No. DCMS03013220).

# REFERENCES

* Electronic address: vnesterenko@ucsd.edu

1.  V.F. Nesterenko; *Dynamics of Heterogeneous Materials, Chapter 1* (Springer-Verlag, NY, 2001).

2.  V.F. Nesterenko; A.N. Lazaridi and E.B. Sibiryakov, *Prikl. Mekh. Tekh. Fiz*. **2**, 19 (1995) [J. Appl. Mech. Tech. Phys. **36**, 166 (1995)].

3.  J. Hong & A. Xu, *Appl. Phys. Lett*., **81**, 4868 (2002).

4.  J. Hong, *Phys. Rev. Lett*. **94**, 108001 (2005).

5.  V.F Nesterenko, C. Daraio, E.B. Herbold, S. Jin, *Phys. Rev. Lett.* in press, see also <arXiv:Cond-mat/0506349> (2005).

6.  L. Vergara, *Phys. Rev. Lett*. **95**, 108002 (2005).

7.  S. Job, F. Melo, , Sen, S. & Sokolow, A. *Phys. Rev. Lett.*, **94**, 178002 (2005).

8.  D.J. Benson, V.F. Nesterenko, *J. Appl. Phys*. **89**, 3622 (2001).

9.  V.F. Nesterenko, "Shock (Blast) Mitigation by "Soft" Condensed Matter" in "Granular Material-Based Technologies", *MRS Symp. Proc*., **729** (MRS, Pittsburgh, PA, 2003), pp. MM4.3.1- 4.3.12.

10. C. Goldenberg & I. Goldhirsch, *Nature*, **435**, 188 (2005).

11. E.I. Corwin,; H.M. Jaeger and S.R. Nagel, *Nature*, **435**, 1075 (2005).

12. T.S. Majmudar and R.P. Behringer, *Nature*, **435**, 1079 (2005).

13. C. Daraio, V. F. Nesterenko, S. Jin, Shock Compression of Condensed Matter-2003, edited by M.D. Furnish, Y.M. Gupta, and J.W. Forbes, American Institute of Physics, AIP Conference Proceedings **706**, 197 (2004).

14. M. Remoissenet,; *Waves Called Solitons (Concepts and Experiments)*, (3rd revised and enlarged edition. Springer-Verlag, Berlin, 1999).

15. C. Coste, E. Falcon & S. Fauve, *Phys. Rev. E*, **56**, 6104 (1997).

16. C. Daraio, V.F.Nesterenko, E.B. Herbold, S. Jin, <arXiv:cond-mat/0506513>, (2005).

17. S. Sen, M.Manciu and J.D. Wright, *Phys. Rev. E*, **57**, 2386 (1998).

18. C. Daraio, V.F. Nesterenko, E.B. Herbold, S. Jin, *Phys. Rev E*, **72**, 016603 (2005).

19. ASM Metals Reference Book, Second Edition, p. 268.

20. ASM Metals Handbook, Properties and Selection of Metals, 8th Edition, **1**, 52 (also www.MatWeb.com).

21. DuPont Product Information, Comparison of Different DuPont Fluoropolymers, www.dupont.com/teflon/chemical/.

22. W.J. Carter and S.P. Marsh, *Hugoniot Equation of State of Polymers*, (Los Alamos Preprint, LA-13006-MS, 1995).

23. S. Jin, T.H. Tiefel, R. Wolfe et. al., *Science*, **255**, 5043, 446 (1992).